\documentclass[prmaterials,twocolumn,superscriptaddress,floatfix,showpacs,amsmath,amssymb]{revtex4}

\usepackage{epstopdf}
\usepackage{graphicx,color}
\usepackage{dcolumn}
\usepackage{bm}
\usepackage{amsmath}
\usepackage[latin1]{inputenc}
\usepackage{epstopdf}
\usepackage{subfigure}
\usepackage{textcomp}
\usepackage{booktabs,multirow}           
\newcommand{\nto}{NiTiO$_3$}
\newcommand{\mto}{MnTiO$_3$}
\newcommand{\cto}{CoTiO$_3$}
\newcommand{\fto}{FeTiO$_3$}
\newcommand{\tn}{$T_{\rm N}$}

\newcommand{\mb}{$\mu_{\rm B}$}
\newcommand{\mbfu}{\(\mu _{\rm B}/{\rm f.u.}\)}

\newcommand{\meff}{$\mu_{\rm eff}$}
\newcommand{\cp}{$c_{\rm p}$}

\newcommand{\cpm}{$c_{\rm p}^{\rm mag}$}
\newcommand{\cpph}{$c_{\rm p}^{\rm ph}$}

\newcommand{\jmk}{J/(mol\,K)}

\newcommand{\ala}{$\alpha_{\rm a}$}
\newcommand{\alc}{$\alpha_{\rm c}$}
\newcommand{\alv}{$\alpha_{\rm V}$}
\newcommand{\alm}{$\alpha_{\rm V}^{\rm mag}$}
\newcommand{\alp}{$\alpha_{\rm V}^{\rm ph}$}

\begin{document}

\title{Magnetic phase diagram and magneto-elastic coupling of \nto}

\author{K.~Dey \footnotemark[2]}
\affiliation{Kirchhoff Institute of Physics, Heidelberg University, INF 227, D-69120 Heidelberg, Germany}
\email[Email:]{kaustav.dey@kip.uni-heidelberg.de}
\author{S.~Sauerland~\footnotemark[2] \footnotetext[2]{Both authors contributed equally.}}
\affiliation{Kirchhoff Institute of Physics, Heidelberg University, INF 227, D-69120 Heidelberg, Germany}
\author{J.~Werner}
\affiliation{Kirchhoff Institute of Physics, Heidelberg University, INF 227, D-69120 Heidelberg, Germany}\author{Y.~Skourski}
\affiliation{Dresden High Magnetic Field Laboratory (HLD-EMFL), Helmholtz-Zentrum Dresden Rossendorf, D-01314 Dresden, Germany}
\author{M.~Abdel-Hafiez}
\affiliation{Harvard University, Cambridge, MA 02138, USA}
\author{R.~Bag}
\affiliation{Indian Institute of Science Education and Research, Pune, Maharashtra 411008, India}
\author{S.~Singh}
\affiliation{Indian Institute of Science Education and Research, Pune, Maharashtra 411008, India}
\author{R.~Klingeler}
\affiliation{Kirchhoff Institute of Physics, Heidelberg University, INF 227, D-69120 Heidelberg, Germany}
\affiliation{Centre for Advanced Materials (CAM), Heidelberg University, INF 225, D-69120 Heidelberg, Germany}



\date{\today}

\begin{abstract}
We report high-resolution dilatometry on high-quality single crystals of \nto\ grown by means of the optical floating-zone technique. The anisotropic magnetic phase diagram is constructed from thermal expansion and magnetostriction studies up to $B=15$~T and magnetisation studies in static (15~T) and pulsed (60~T) magnetic fields. Our data allow to quantitatively study magneto-elastic coupling and to determine uniaxial pressure dependencies. While the entropy changes are found to be of magnetic nature, Gr\"{u}neisen analysis implies only one relevant energy scale in the whole low temperature regime. Thereby, our data suggest that the observed structural changes due to magneto-elastic coupling and previously reported magnetodielectric coupling~\cite{Harada} are driven by the same $magnetic$ degrees of freedom that lead to long-range magnetic order in \nto , which, in turn, establishes a linear magnetodielectric coupling in this compound.
\end{abstract}

\maketitle

\section{Introduction}

The search for new multiferroics, i.e., materials concomitantly exhibiting various ferroic orders such as magnetic and electric order coupled to each other~\cite{Spaldin391}, has revived interest in the family of ilmenite-structured compounds. There are various titanates $M$TiO$_3$ ($M$ = Mn, Fe, Co, Ni) which crystallize in the ilmenite structure within the space group $R$-3. The crystal structure consists of alternate layers of corner sharing TiO$_6$ and $M$O$_6$ octahedra stacked along the $c$-axis~\cite{barth}. Previous magnetic~\cite{Stickler,Watanabe, Heller} and powder neutron diffraction studies~\cite{Shirane, Newnham, YAMAGUCHI,Charilaou} report that long range antiferromagnetic (AFM) order evolving at low temperatures is of the G-type for \mto , and of the A-type for \fto, \cto\ and \nto. In the ordered phase, the magnetic moments associated with $M^{2+}$-ions are collinearly arranged along the $c$-axis in \mto\ and with a slight spin tilting of 1.6\,° away from the $c$-axis in \fto. The easy-plane-type AFM order in \cto\ and \nto\ is characterised by spins lying in the ferromagnetic $ab$-layers which are aligned antiferromagnetically along $c$. In \fto, the onset of long range magnetic order, at \tn , is associated with significant changes of the lattice parameters indicating magneto-elastic coupling~\cite{Charilaou}. Magnetodielectric and polarization measurements on \mto\ show an anomaly in the dielectric permittivity, $\epsilon$, at \tn , while finite polarisation is found in applied external magnetic fields indicating that it may realise a linear magnetoelectric material~\cite{Mufti}. Both in \cto\ and \nto, anomalies in $\epsilon$ at \tn\ and a strong field dependant magnetocapacitance in the ordered state indicate the presence of large magneto-dielectric coupling~\cite{Harada}.

Despite clear evidence of pronounced magneto-dielectric coupling in all known $M$TiO$_3$, its origin and mechanism have not yet been elucidated. In order to address this question, we have grown sizeable, high-quality single crystals of \nto\ by means of the optical floating-zone technique under various atmospheres and at different pressure. The single crystals were used for high-resolution studies of thermal expansion and magnetostriction along the crystallographic $a$- and $c$-axis, respectively. In addition, comparing the magnetic length and entropy changes as detected by thermal expansion coefficients and specific heat allow determining the uniaxial pressure dependencies by means of Gr\"{u}neisen scaling. Moreover, these results as well as magnetisation studies in static (15~T) and pulsed (60~T) magnetic fields give access to the anisotropic phase diagram in \nto .

\section{Experimental methods}

\nto\ powder was prepared via standard solid-state reaction of stoichiometric amounts of NiO and TiO$_2$ between 1150 and 1350\,°C with several intermediate grinding steps. The powder was made into rods of length 10~cm and 5~mm in diameter by hydrostatically pressing the powders at 700~bar and annealing them for 24~h at 1350\,°C. Single crystals of \nto\ were grown in a four-mirror optical floating-zone furnace (Crystal system corporation, Japan) equipped with $4\times150$~W halogen lamps at IISER Pune and in a two mirror high-pressure optical floating-zone furnace (HKZ, SciDre) equipped with a 3500~W Xe arc lamp at Heidelberg University. Macroscopic single crystals were grown at 3~mm/h under various atmospheres and up to 5~bar pressure. Phase purity of the powder and the ground single crystals was studied by means of powder X-ray diffraction measurements on a Bruker D8 Advance ECO diffractometer with Cu-K$\alpha$ source. Laue diffraction in back scattering geometry was performed to study the crystallinity and to orient the single crystals. Structural Rietveld refinements were carried out using the Full Prof suite 2.0\cite{fullprof}.

Static magnetisation $\chi=M/B$ was studied in magnetic fields up to 15~T applied along the principal crystallographic axes by means of a home-built vibrating sample magnetometer~\cite{vsm} (VSM) and in fields up to 5~T in a Quantum Design MPMS-XL5 SQUID magnetometer. Pulsed-magnetic-field magnetization was studied up to 60~T at Helmholtz Zentrum Dresden Rossendorf by an induction method using a coaxial pick-up coil system~\cite{Pulsed}. The pulse raising time was 7~ms. The pulsed-field magnetization data were calibrated using static magnetic field measurements. Specific heat measurements at 0~T and 9~T has been done in a Quantum Design PPMS using a relaxation method. The relative length changes $dL_i/L_i$ were studied on a cuboid shaped single crystal of dimensions $2 \times 1.85 \times 1~$mm$^{3}$. The measurements were done by means of a three-terminal high-resolution capacitance dilatometer~\cite{dil}. In order to investigate the effect of magnetic fields, the linear thermal expansion coefficient $\alpha_i = 1/L_i\cdot dL_i(T)/dT$ was studied in magnetic fields up to 15~T. In addition, the field induced length changes $dL_i(B)/L_i$ were measured at various fixed temperatures in magnetic fields up to 15~T and the longitudinal magnetostriction coefficient $\lambda_i = 1/L_i\cdot dL_i(B)/dB$ was derived. The magnetic field was applied along the direction of the measured length changes.

\section{\nto\ single crystal growth}

\begin{figure}[h]
\centering
\includegraphics [width=0.95\columnwidth,clip] {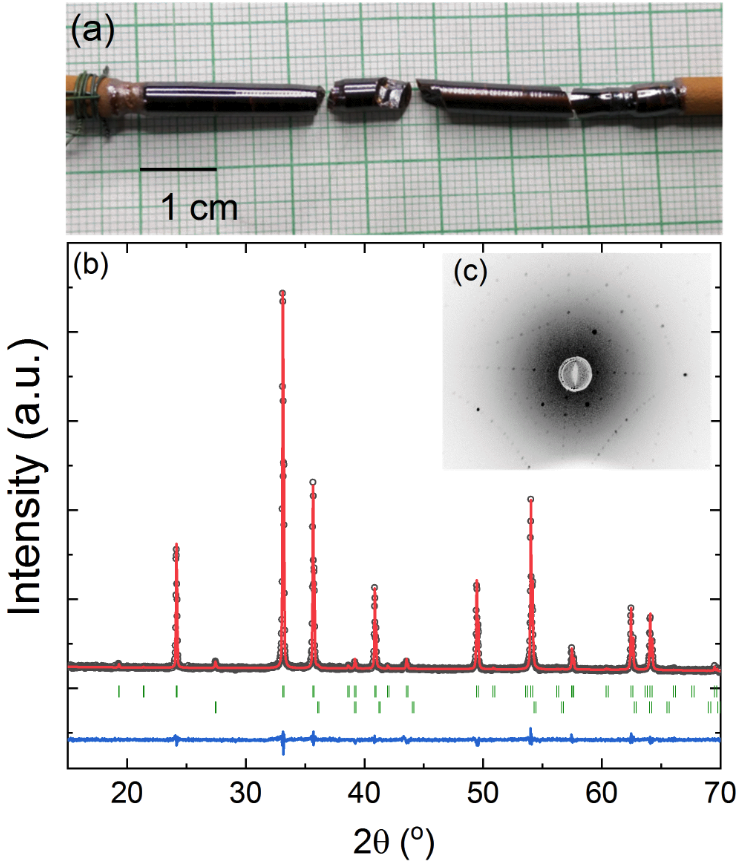}
\caption{(a) Picture of a shiny \nto\ boule grown in air atmosphere at ambient pressure, (b) Rietveld refinement fit of the room temperature XRD data of a powdered \nto\ single crystal. The observed diffraction pattern is shown in black, calculated pattern in red and the difference between the observed and the calculated pattern is shown in blue. The upper vertical ticks in green denote the allowed Bragg positions of the ilmenite phase and the lower ticks denote the Bragg positions of TiO$_2$ in rutile phase. (c) Representative Laue pattern of the \nto\ single crystal oriented along (010) direction.} \label{crystal}
\end{figure}

Single crystals of \nto\ were grown by the optical floating-zone method using polycrstalline feed and seed rods as starting material. The phase purity of the \nto\ powders used for making the feed and seed rods has been studied by means of powder XRD after each sintering step. Rietveld refinement of these data indicates the presence of ilmenite phase ($R$-3) as well as an additional ($\approx$1\%) TiO$_2$ (rutile) phase. In order to achieve phase pure high-quality single crystals, a variety of crystal growth experiments were performed under different atmosphere and pressure and by means of both the four-mirror (CSC, horizontal configuration) 
and two-mirror (HKZ, vertical configuration)~\cite{HERGETTLFSO,NEEF2017,WIZENT2009,WIZENT2011995,WIZENT2014} high-pressure optical floating-zone furnaces. The optimized growth parameters employed during crystal growth are listed in Table \ref{table1}.

\begin{table}[htb]
\centering
\caption{Growth parameters, lattice parameters and phase analysis from the Rietveld refinement of the room temperature powder XRD data of crushed \nto\ single crystals. Feed and seed rods were counter-rotated at the same rotation speed.}
\begin{tabular}{l|ccc}
furnace& CSC & CSC & HKZ \\
\hline\hline
atmosphere&O$_2$&air &Ar\\
pressure&ambient&ambient&5~bar\\
growth rate (rpm)&3 &3 &4-6\\
rotation speed (mm/h)&20 &10 &15\\
latt. parameter $a$ (\AA)&5.0304 &5.0304 &5.0304\\
latt. parameter $c$ (\AA)&13.7881 &13.7845 &13.7862\\
\hline
crystal size&cm&cm&mm\\		
secondary phase (appr.)&1\% TiO$_2$+NiO& 1\% TiO$_2$& 1\% NiO \\		
\end{tabular}
\label{table1}
\end{table}

Depending on the growth parameters, the processes summarized in Table~\ref{table1} yield mm- to cm-sized single crystals. Fig.~\ref{crystal}a shows a representative \nto\ boule grown in air at ambient pressure. The shiny surface of the grown boule indicating the presence of single crystal of several cm in length. Rietveld refinement of powder XRD data of the ground and pulverized single crystalline pieces extracted from the boules (see Fig.~\ref{crystal}b and Table~\ref{table1}) implies the main ilmenite phase as well as an impurity TiO$_2$ phase of about 1~\%. When grown in O$_2$ atmosphere, there is also an additional NiO phase. 
In order to further investigate the growth process of \nto\ and to reduce the secondary phase content, \nto\ was also grown under 5~bar pressure in Ar atmosphere. In this case, the resulting boule is mostly of polycrystalline nature with only mm-sized shiny single crystalline regions towards the end of the boule. Phase analysis of a single crystalline piece extracted from this region shows the presence of about 1~\% NiO secondary phase while the TiO$_2$ phase is absent. We conclude that inert atmosphere does not support optical floating-zone-growth of \nto\ single crystals. For the magnetic studies presented below, we employ crystals grown in air as they exhibit only a small non-magnetic impurity phase. Laue diffraction performed at several spots along the length of grown boules  confirm the presence of macroscopic cm-sized single crystalline grains with high crystallinity (see Fig.~\ref{crystal}c). For the measurements reported below, a cuboid shaped single crystal of dimensions $2 \times 1.85 \times 1~$mm$^{3}$ has been extracted from the boule grown in air and oriented along three principal crystallographic directions.

\section{Experimental results}

\subsection{Magneto-elastic coupling}

\begin{figure}[tb]
\centering
\includegraphics [width=0.95\columnwidth,clip] {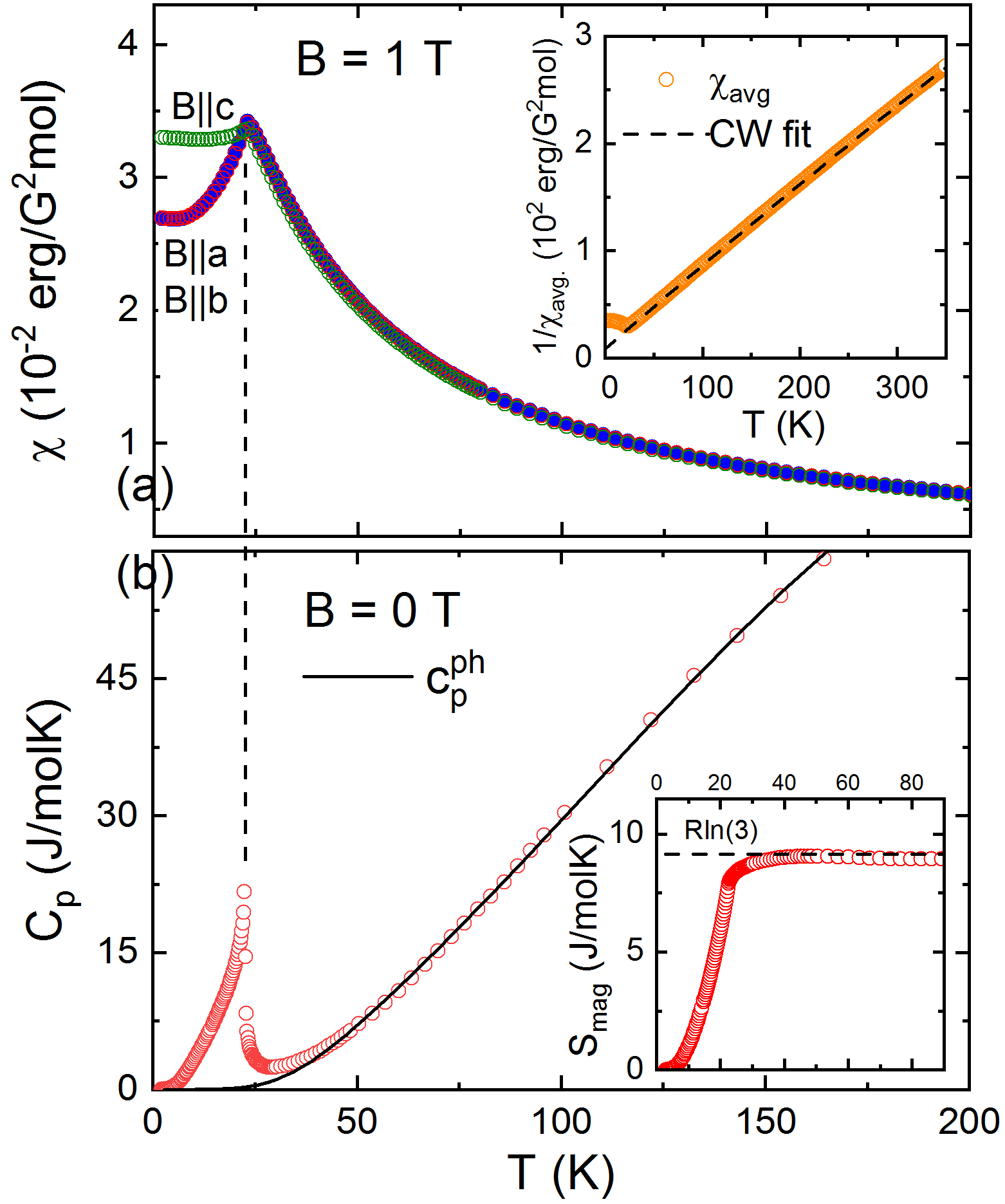}
\caption{(a) Temperature dependence of the static magnetic susceptibility $\chi =M/B$ at $B=1$~T applied along the main crystallographic directions and (b) temperature dependence of the specific heat \cp\ in zero magnetic field. The solid line in (b) indicates the phonon specific heat \cpph\ obtained by fitting \cp\ data with a combined Debye- and Einstein-model well above the magnetic ordering transition (see the text). Insets: (a) Curie-Weiss fit (dashed line) to the inverse averaged susceptibility $1/\chi_{av}$. (b) Magnetic entropy changes obtained by integrating (\cp -\cpph )/$T$.} \label{chi}
\end{figure}

The onset of long range antiferromagnetic order in \nto\ at \tn\ = 22.5(5)~K is associated with pronounced anomalies in magnetic susceptibility and specific heat (Fig.~\ref{chi}). For $T\leq T_{\rm N}$, the susceptibility is anisotropic with a decrease for magnetic fields $B$ applied in the $ab$-plane and attaining a constant value for $B||c$-axis suggesting an easy-plane-type antiferromagnet. This is in accordance with the previous studies~\cite{Shirane,Watanabe}. At high temperatures, the susceptibility is isotropic and obeys a Curie-Weiss behaviour. Fitting the averaged susceptibility (inset to Fig.~\ref{chi}a) at $T>100$~K by means of a Curie-Weiss-like law $\chi = \chi_{0}+N_{\rm A} \mu_{\rm eff}^2/3k_{\rm B}(T-\Theta)$ yields $\chi_{0}=1.93\times10^{-4}$, the effective magnetic moment \meff\ $= 3.17(5)$~\mb\ and the Weiss temperature $\Theta=-11(1)$~K. Using the spin-only value $S = 1$ for Ni$^{2+}$, implies an effective $g$-factor of $2.24(4)$. Note, that our measurements yield a smaller value than \meff\ = $4.01$~\mb\ previously reported for a single crystal~\cite{Watanabe} but is similar to the values reported for polycrystalline samples~\cite{Stickler,Harada}.

The sharp $\lambda$-shaped anomaly in the specific heat (Fig.~\ref{chi}b) confirms the onset of long-range magnetic order at \tn\ and also indicates high crystallinity of the single crystal. Furthermore, the anomaly presents a continuous nature of the phase transition. The phonon contribution to the specific heat (\cpph ) has been estimated by fitting the \cp\ data at temperatures well above \tn\ by an extended Debye-model which includes both Debye- and Einstein-terms~\cite{book}. The model fits very well for temperatures above about 50~K and yields characteristic Debye- and Einstein-temperatures of $\Theta_D =786$~K and $\Theta_E =230$~K, respectively. The sum of the obtained coefficients of the individual terms $n_D = 3.94$ and $n_E = 0.94$ reasonably agrees to the expected value of 5 which reflects the number of phonon modes given by the number of atoms per formula unit. Integrating the magnetic specific heat (\cp -\cpph )/$T$ yields a total magnetic entropy change $S_{\rm mag} = 9.1(1)$~\jmk\ which agrees to the theoretically expected value for $S=1$ Ni$^{2+}$ spins of $R\ln (3)= 9.13$~\jmk. We conclude that the entropy changes are of magnetic nature. The measured entropy changes imply that nearly 20\% of magnetic entropy is consumed between \tn\ and 50~K, suggesting the presence of short-range magnetic correlation persisting up to temperatures as high as twice the ordering temperature.

\begin{figure}[htb]	
\includegraphics[width=0.95\columnwidth,clip] {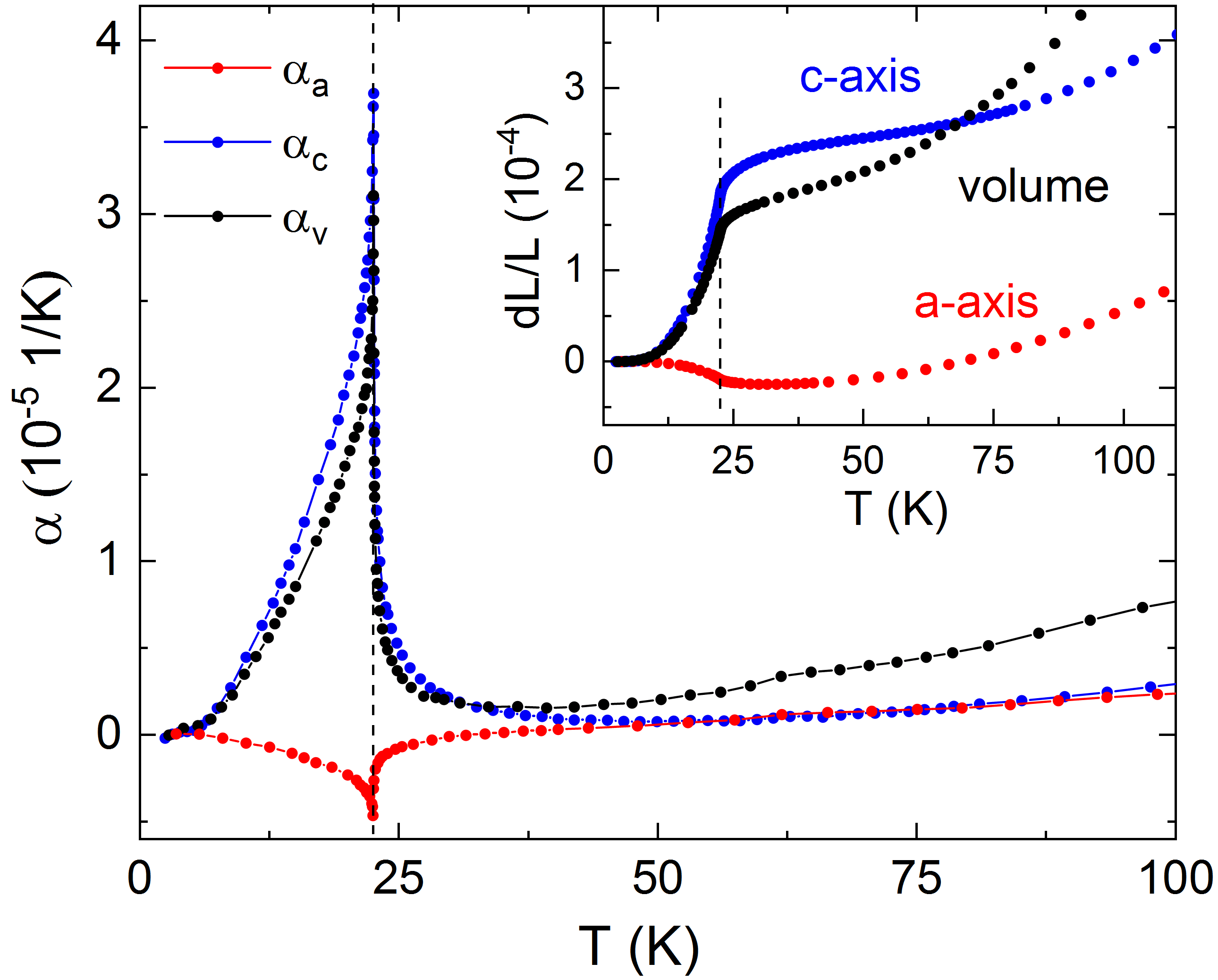}
\caption{Thermal expansion coefficients $\alpha_i$ along the crystallographic $a$- and $c$-axes and the volume thermal expansion coefficient $\alpha_{\rm V}$. The inset shows the relative length changes $d L_i/L_i$. Dashed lines mark \tn .}\label{alpha}
\end{figure}

The evolution of long range magnetic order is associated with pronounced length changes as illustrated by strong anomalies in the uniaxial thermal expansion coefficients $\alpha_{\rm i}$ ($i=a,c$) and in the relative length changes $dL_i/L_i$ (Fig.~\ref{alpha}). The anomalies demonstrate the presence of significant magneto-elastic coupling in \nto. The measured relative length changes shown in the inset of Fig.~\ref{alpha} signal shrinking of the $c$-axis and increase of the $a$-axis upon evolution of magnetic order at \tn . Thus, the signs of the anomalies indicate positive uniaxial pressure dependence of \tn\ for pressure along $c$-axis, i.e $\partial T_{\rm N}/\operatorname{\partial p_c}>0$, whereas the anomaly in $\alpha_{\rm a}$ indicates $\partial T_{\rm N}/\operatorname{\partial p_a}$ being negative and considerably smaller. The anomaly in the volume thermal expansion coefficient $\alpha_{\rm V}=\alpha_{\rm c}+2\alpha_{\rm a}$ implies a significant positive $hydrostatic$ pressure dependency of \tn . In addition, opposite sign of the anomalies in \ala\ and \alc\ allows to conclude that effects in the thermal expansion associated with magnetic order extend up to about 50~K which agrees to the temperature regime where magnetic entropy changes mark the onset of short-range magnetic correlations.

\subsection{Magnetic phase diagram}

\begin{figure}[htb]
\centering
\includegraphics [width=0.95\columnwidth,clip] {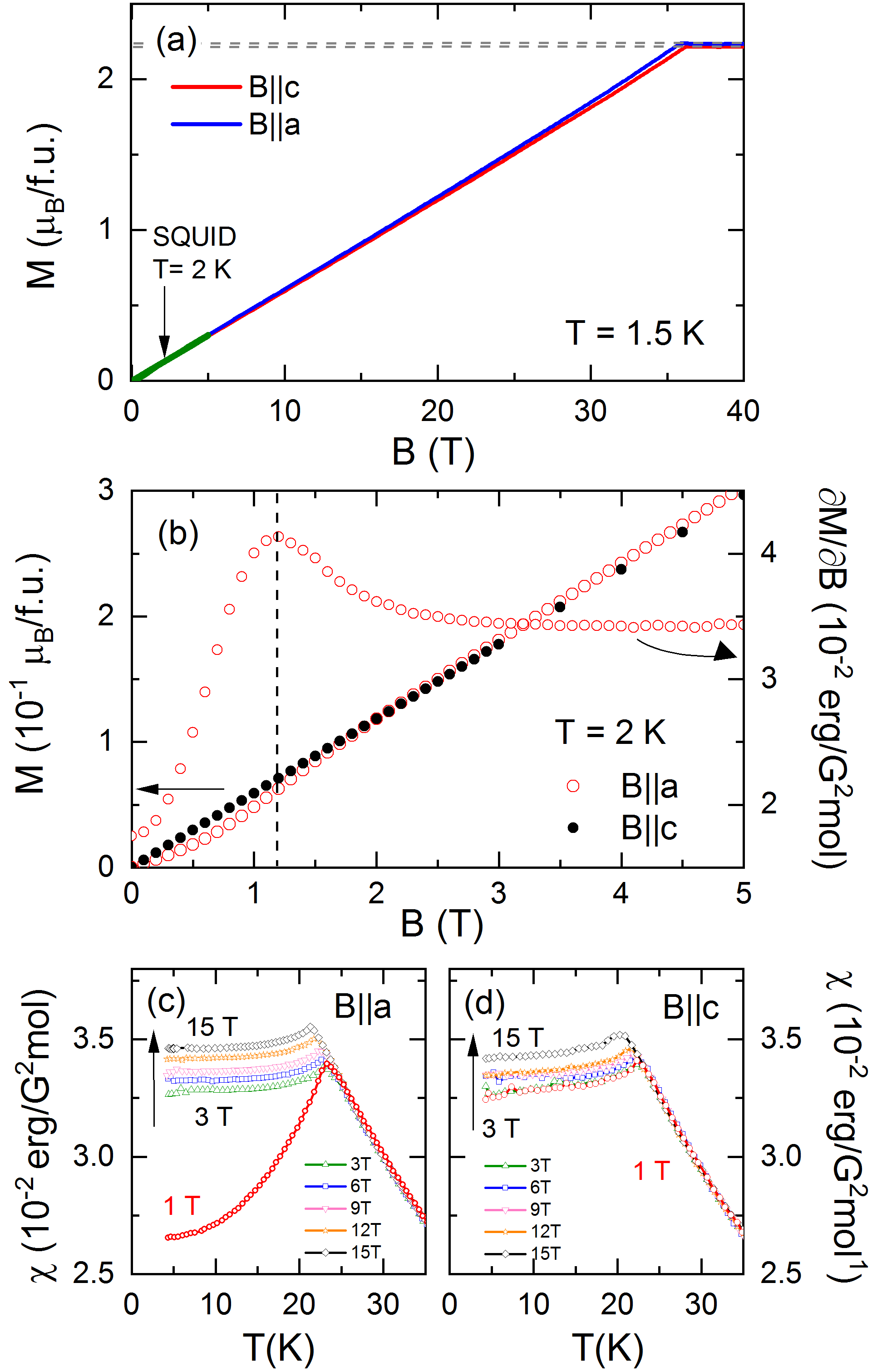}
\caption{(a) Pulsed-field magnetization $M$ at $T=1.5$~K, (b) quasi-static field magnetisation $M$ and magnetic susceptibility $\partial M/\partial B$ vs. magnetic field along the $a$- and $c$-axes, at $T=2$~K, (c) and (d) static magnetic susceptibility $\chi = M/B$ vs. temperature for magnetic fields up to 15~T applied along the $a$- and $c$-axes, respectively.} \label{mag}
\end{figure}

The saturation fields and moments at $T = 1.5~K$ are deduced from pulsed-field magnetisation studies up to 60~T which are shown in Fig.~\ref{mag}a. For both field directions $B||c$ and $B||a$, the magnetisation shows a linear behaviour in a wide range of applied fields. The saturation fields nearly coincide and amount to $B_{\rm sat} = 36.0(5)$~T. Also the saturation magnetisations as indicated by the dashed horizontal lines in Fig.~\ref{mag}a agree within error bars at $M_{\rm sat} = 2.23(5)$~\mbfu. Assuming the spin-only momentum $S = 1$ yields a $g$-factor of $2.23(5)$ which agrees well with the value of $2.24(4)$ derived from the Curie-Weiss fit to the static magnetic susceptibility presented in Fig.~\ref{chi}. A more detailed look at the low-field behaviour in Fig.~\ref{mag}b, at $T=2$~K, confirms that linearity of $M$ vs. $B||c$ extends to zero magnetic field while non-linear behaviour is observed when the magnetic field is applied along the $a$-axis. To be specific, the derivative of the magnetization with respect to magnetic field shows a broad peak centered at $B^* = 1.20(5)$~T and subsequently a constant behaviour (see Fig.~\ref{mag}b). The data suggest spin reorientation which we attribute to finite anisotropy in the $ab$-plane. At $T=2$~K, the magnetisation jump at $B^*$ is estimated to $\Delta M\approx 0.03$~\mb /f.u. Remaining slight non-linearity above $B^*$ is indicated by the static magnetic susceptibility measured in magnetic fields up to 15~T (Fig.~\ref{mag}c and d). While there is no significant field effect for $T >$ \tn , the data exhibit a monotonous change for $T <$ \tn\ at applied fields $B \geq 3$~T as compared to $B = 1$~T. Overall, the data confirm spin-reorientation behaviour as for $B \geq 3$~T (i.e., above $B^*$) the susceptibility attains an almost constant value below \tn\ whereas it decreases sharply for $B||a = 1$~T. In addition, the phase boundary \tn ($B$) marked as peak in the static susceptibility is determined.

\begin{figure}[tb]
\includegraphics[width=0.95\columnwidth]{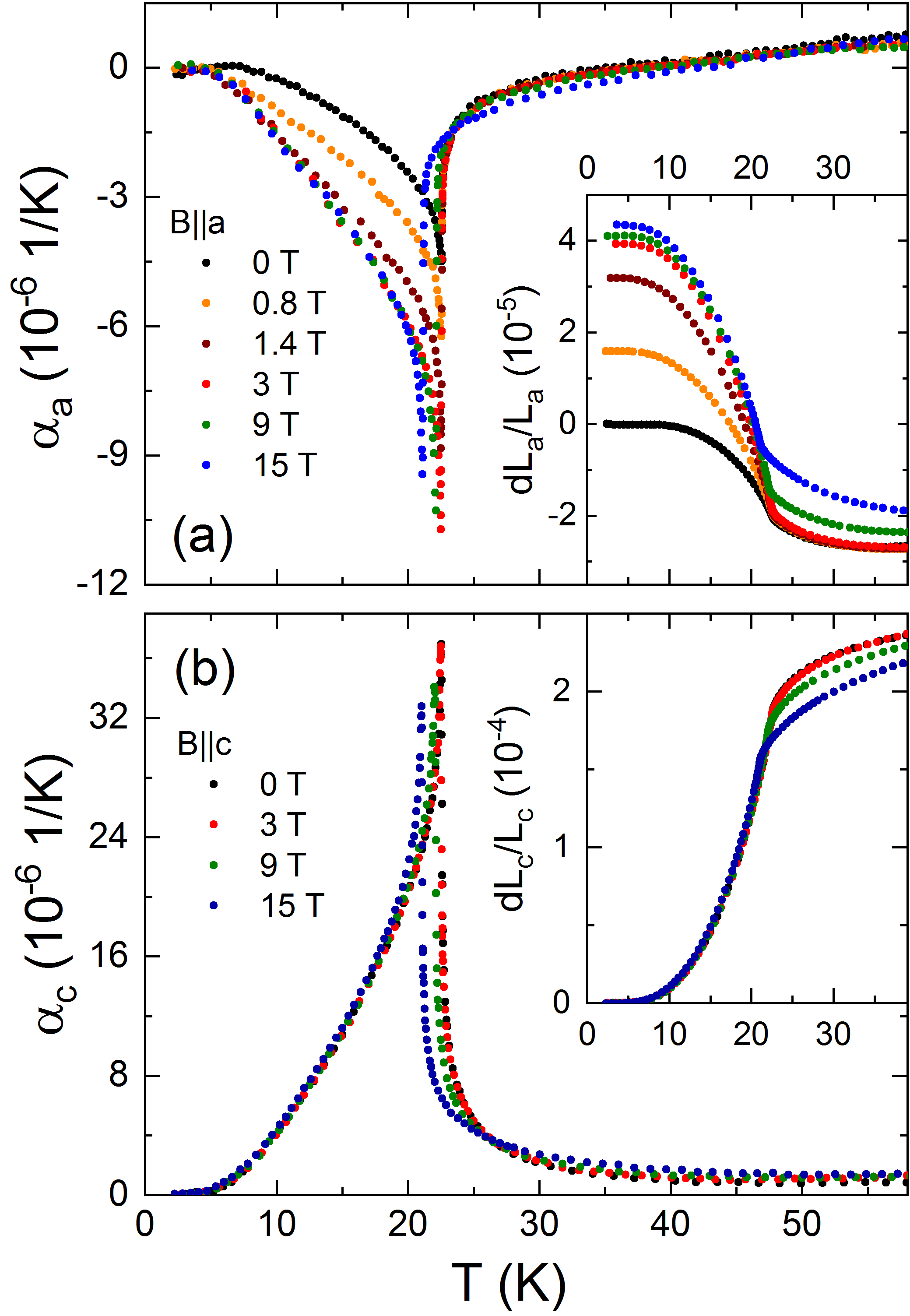}
\caption{Thermal expansion coefficients $\alpha_{\rm i}$ at magnetic fields between 0 T and 15 T applied along the $a$- and $c$-axes, respectively, of \nto. Insets show the corresponding relative length changes shifted with respect to each other by means of magnetostriction curves, at $T=30$~K.}\label{alpham}
\end{figure}

\begin{figure}[htb]	
\includegraphics[width=0.95\columnwidth,clip] {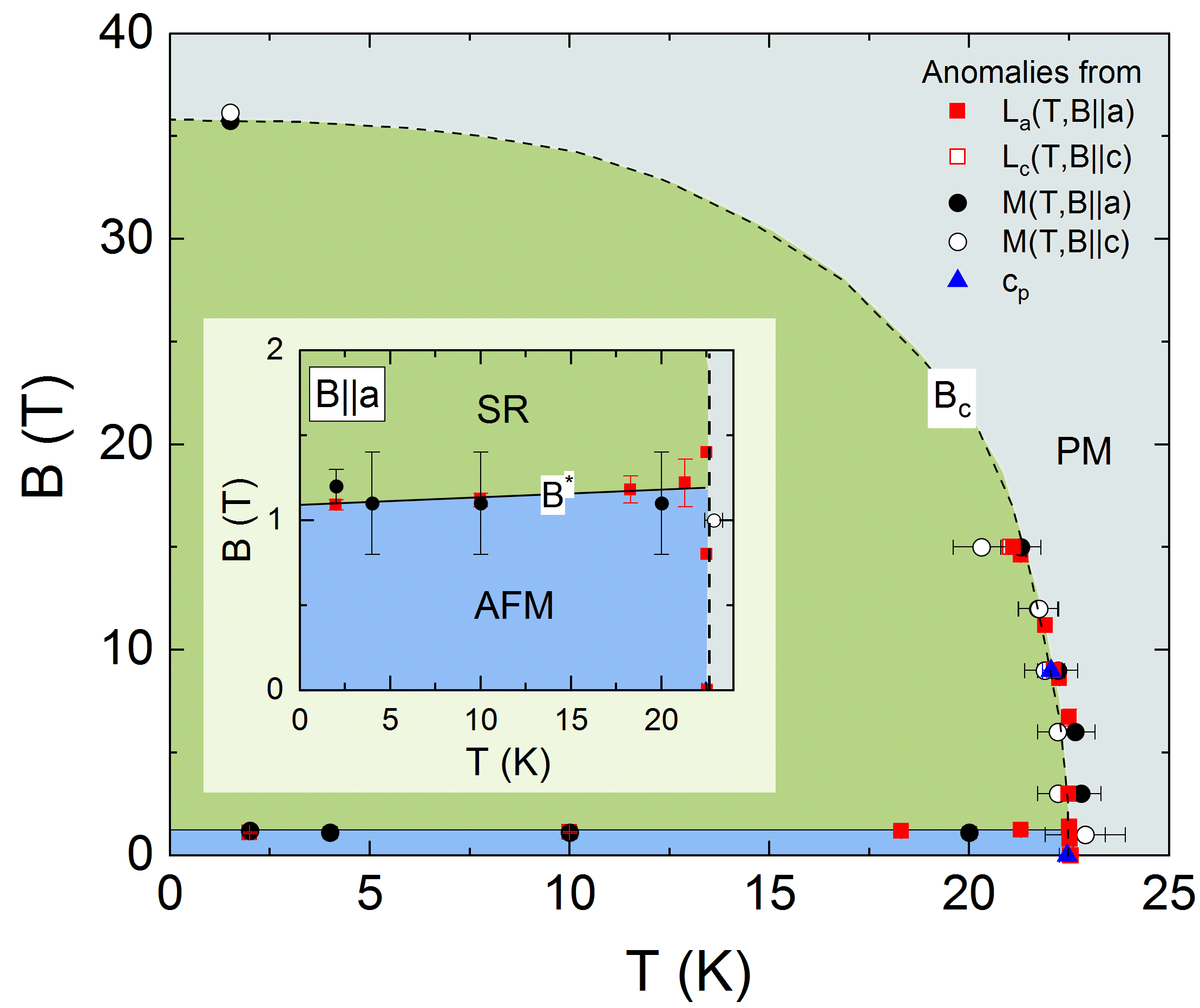}
\caption{Magnetic phase diagram of \nto\ constructed from magnetisation $M$($T$,$B$), dilatometry $L$($T$,$B$) and specific heat data. Closed (open) markers correspond to magnetic fields applied along the $a$-axis ($c$-axis). Lines are guides to the eye. AFM, SR, PM label the antiferromagnetically ordered, spin-reoriented and paramagnetic phases respectively.} \label{phd}
\end{figure}

Sharp lambda-shaped anomalies observed in $\alpha_{\rm i}$ ($i=a,c$) in external magnetic fields (Fig.~\ref{alpham}) enable to further determine the phase boundaries and to study the magnetic field effect on the lattice parameters. While the shape of the anomalies is not significantly affected by magnetic fields, \tn\ expectedly shifts to lower temperatures upon application of external magnetic fields. For both field directions, a similar shift of $\Delta T_{\rm N}\approx 1.5$~K is observed when applying $B=15$~T. This corroborates well with the magnetization data in Fig.~\ref{mag}c,d and signals overlying phase boundaries for $B$ applied along the $a$- and $c$-axis, respectively (see Fig.~\ref{phd}). Corresponding anomalies signalling \tn ($B$) (or synonymously the temperature dependence of the critical field $B_c(T)$ which signals melting of magnetic order, too) appear in the relative length changes versus magnetic field (Fig.~\ref{msall}, see below) and in the magnetostriction coefficients (see supplement Fig.~S1) and enable constructing the magnetic phase diagram displayed in Fig.~\ref{phd}. Up to 15~T, \tn ($B$) obeys a square-root behaviour. 

\begin{figure}[htb]	
\includegraphics[width=.95\columnwidth,clip] {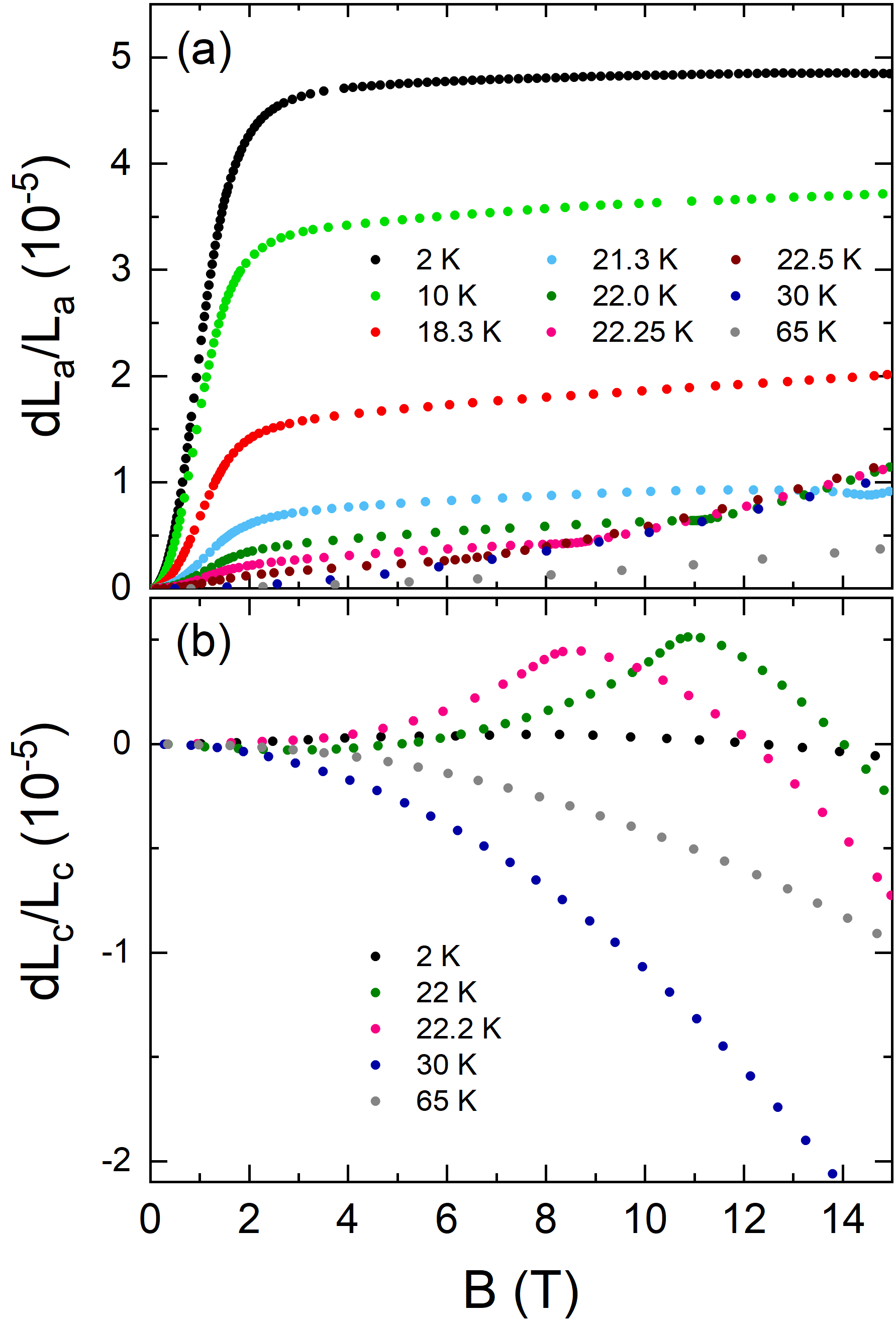}
\caption{Relative length changes versus magnetic field applied along $a$- and $c$-axes at different temperatures.}\label{msall}
\end{figure}

The thermal expansion in magnetic field and the magnetostriction data in Figs.~\ref{alpham} and \ref{msall}, respectively, show small increase of the $a$-axis and decrease of the $c$-axis in magnetic fields applied in the respective directions. Interestingly, except for effects associated with suppression of \tn , there are no large length changes $dL_c$ driven by $B\|c\leq 15$~T. Correspondingly, the magnetostriction $\lambda_c$, at 2~K (see supplement Fig.~S1), is small and amounts to a few 10$^{-7}$~T$^{-1}$ only. While a similar behaviour is found for $\lambda_a$ at $B\|a\gtrsim 5$~T, there are pronounced length changes at low fields which we associate with spin-reorientation. The corresponding half height of these jump-like anomalies in $dL_a/L_a$ is at consistent fields as the peak in $\partial M/\partial B$ and yields $B^*$ (see Fig.~\ref{phd}). Notably, these field-induced changes imply that the total thermal expansion changes in the magnetically ordered phase become considerably larger in applied magnetic fields (see the inset of Fig.~\ref{alpham}a). Quantitatively, spin-reorientation at 2~K is associated with length changes of $\Delta L_a/L_a \approx 4.8\cdot 10^{-5}$. Upon heating, the size of magnetostriction decreases but, nonetheless, significantly exceeds length changes observed at $T>$\tn .

\subsection{Discussion}

\begin{figure}[htb]	
\includegraphics[width=0.95\columnwidth,clip] {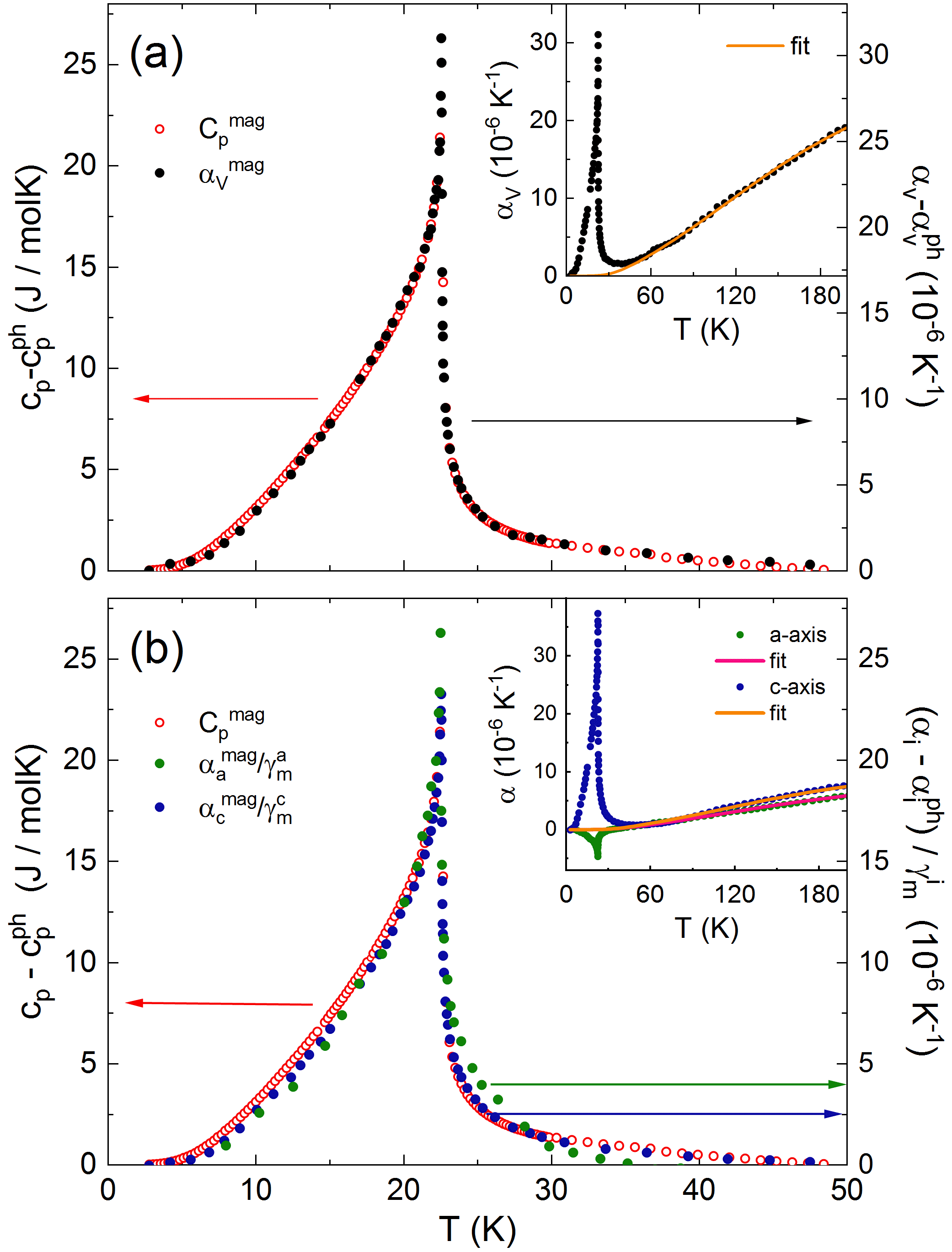}
\caption{Gr\"{u}neisen scaling of the magnetic contributions to the heat capacity (\cpm ) and (a) volume thermal expansion coefficient (\alm ). The inset shows $\alpha_{\rm V}$ vs. temperature together with a combined Debye-Einstein fit to the high-temperature data (see the text). (b) Gr\"{u}neisen scaling with uniaxial thermal expansion coefficients \ala\ and \alc .}\label{scaling}
\end{figure}

Comparing the non-phononic contributions to the thermal expansion coefficient and to the specific heat enables further conclusions on the nature of the associated (i.e., magnetic) entropy changes and on the hydrostatic pressure dependencies. In order to assess the magnetic contribution to the volume thermal expansion coefficient, \alm , we have approximated the phonon contribution, \alp , by scaling the background specific heat \cpph\ (cf. Fig.~\ref{chi}) appropriately. For that purpose, a Debye-Einstein model with fixed $\Theta_D$ and $\Theta_E$ of the fit to the specific heat data has been fitted to the high-temperature data of \alv\ leaving only the two associated lattice Gr\"{u}neisen parameters $\gamma_D$ and $\gamma_E$ as scaling factors. Similar to the specific heat, at $T\gtrsim 50$~K, \alv\ is well described by the phonon background \alp\ with $\gamma_{D} = 2.8\times 10^{-7}$~mol/J and $\gamma_{E} = 2.3\times 10^{-7}$~mol/J as shown in the inset of Fig.~\ref{scaling}.

The resulting non-phonon contribution \alm\ to the thermal expansion coefficient is shown in Fig.~\ref{scaling}a (right ordinate) together with the respective magnetic specific heat data \cpm\ (left ordinate). Both quantities are proportional to each other in the entire temperature range. In the sense of a Gr\"{u}neisen analysis, such behaviour implies the presence of a single dominant degree of freedom from which we conclude that a single magnetic degree of freedom drives the observed non-phonon length and entropy changes. The corresponding scaling parameter obtained is the magnetic Gr\"{u}neisen parameter $\gamma_m =$ \alm / \cpm $=1.18(3)\times10^{-6}$~mol/J. Using the Ehrenfest relation, the obtained value of $\gamma_m$ enables to determine the hydrostatic pressure dependence $dT_{\rm N}/dp = T_{\rm N} V_{m}\gamma_{m} = 1.12(4) $~K/GPa which is deduced using the molar volume of $V_{m} = 42.01$~cm$^{3}$/mol. Elaborating Gr\"{u}neisen scaling for the uniaxial thermal expansion coefficients individually, good proportionality is confirmed between \cpm\ and $\alpha_{\rm a}^{\rm mag}$ and $\alpha_{\rm c}^{\rm mag}$ as well (Fig.~\ref{scaling}b). This yields the uniaxial pressure dependencies of $dT_{\rm N}/dp_a = -0.21(3) $~K/GPa and $dT_{\rm N}/dp_c = 1.51(7) $~K/GPa for pressure applied along the $a$- and $c$-axis, respectively. These values are fully consistent with the obtained hydrostatic pressure dependence.

Magnetostriction measurements in the paramagnetic regime, i.e., where $M=\chi B$, enable to extract the uniaxial pressure dependence of magnetic susceptibility by exploiting the relation $dL_i/L_i = - 1/2V \partial \chi_i / \partial p_i B^2$~\cite{johannsen}. Plotting the data accordingly (see supplement Fig.~S2) allows to read off $\partial (\ln \chi_a) / \partial p_a = -1.3$~\%~GPa$^{-1}$ and $-0.8$~\%~GPa$^{-1}$ at 30 and 65~K, respectively, as well as $\partial (\ln \chi_c) / \partial p_c = +3.1$~\%~GPa$^{-1}$ and $+2.2$~\%~GPa$^{-1}$ at respective temperatures. Qualitatively, this suggests AFM exchange interactions to be strengthened and FM ones to be weakened by uniaxial pressure along the $a$-axis. While, uniaxial pressure along the $c$-axis is found to result in opposite effects. Considering the results of Gr\"{u}neisen analysis presented above, i.e., predominance of only one energy scale as well as $\partial T_N / \partial p_a < 0$ and $\partial T_N / \partial p_c > 0$, suggests that the value of \tn\ is mostly affected by the (in-plane) FM exchange interactions. This is also reflected in the increase of \tn\ and Weiss temperature $\theta$ when substituting Ni over Co to Fe in $M$TiO$_3$ whereby even $\theta>0$ is observed for \fto~\cite{Stickler}.

The phase boundary of spin-reorientation features a very small slope $\partial B^*/\partial T\approx 4\times10^{-3}$~T/K. Considering the magnetisation jump $\Delta M$ at $B^*$ and exploiting the Clausius-Clapeyron relation, we estimate associated entropy changes $\Delta S^*=-(\partial T/\partial B^*)^{-1}\cdot \Delta M^* \approx -8\times10^{-4}$~\jmk ~\cite{stockert}. This implies only insignificant entropy changes associated with spin reorientation. On the other hand, in applied magnetic fields $B\|a$, our data show that the total thermal expansion changes $dL_a/L_a$ in the magnetically ordered phase become significantly larger (see the inset of Fig.~\ref{alpham}). This suggests that Gr\"{u}neisen scaling which is valid at $B=0$~T as evidenced by Fig.~\ref{scaling} is broken in magnetic fields $B||a > B^*$. Somehow correspondingly, uniaxial pressure dependence of $B^*$ is very large. Using the measured jumps in relative length changes ($\Delta L_a/L_a$) and magnetization ($\Delta M$) at $B^*$ and 2~K and exploiting the Clausius-Clapeyron relation yields  $\partial B^*/\partial p_a = V\cdot(\Delta L_a/L_a)/\Delta M \approx 9.2$~T/GPa. This is a huge value similar to what has been observed, i.e., in TlCuCl$_3$~\cite{johannsen}. It implies strong effects of uniaxial pressure along the $a$-axis so that applying $p_a$ would strongly enhance the spin reorientation field while it would vanish for tiny hypothetical negative pressure.

Above the \tn , i.e., in the absence of long-range spin order where short-range correlations are still present as, e.g., indicated by the specific heat data, magnetostriction is relatively large in \nto . This observation agrees to the fact that both $\lambda_a$ and $\lambda_c$ become significantly larger when $B$ exceeds $B_c$ which appears at 21~K $\leq T\leq$ 22.2~K in the accessible field range (see Fig.~\ref{msall} and supplement Fig.~S1). We conclude that, in a paramagnetic but yet correlated regime, magnetic fields along the $a$- and the $c$-axis, respectively, yield reorientation of spins which are short-range ordered in this temperature and field ranges.

Recently, anomalies in the electrical permittivity $\epsilon$ at \tn\ and strongly field-dependant magnetocapacitance close to \tn\ have been observed in polycrystalline \nto\ indicating the presence of significant magneto-dielectric coupling~\cite{Harada}. The shape of the reported~\cite{Harada} temperature dependence of $\epsilon$ is very similar to the length and volume changes observed by our thermal expansion measurements (see supplement Fig.~S3) indicating an almost linear relation between electrical permittivity and structural distortion below \tn . Conclusively, $\epsilon(T)$ and reported magneto-dielectric coupling are directly related to the length changes and the magneto-elastic coupling. Furthermore, driving entropy changes of the low-temperature effects are purely of magnetic nature as evidenced by Gr\"{u}neisen analysis presented above. We conclude that magneto-dielectric coupling is secondarily mediated via structural changes and that magnetic degrees of freedom constitute a single common origin for the dielectric, structural and magnetic changes evolving at and below \tn\ in \nto. Note, however, that magnetocapacitance data on polycrystals (at $T=15$\,K) do not show anomalies at $B^*$ though spin-reorientation is associated with significant length changes. One might speculate that the polycrystalline nature of samples studied in Ref.~\onlinecite{Harada} masks such effects.\\

\section{Summary}

In summary, we report growth and characterization of large and high quality \nto\ single crystals by means of the optical floating-zone technique. The anisotropic phase diagram is constructed by means of pulsed and static magnetization, specific heat, thermal expansion, and magnetostriction data. It features a spin-reorientation transition at $B^* ||a \approx 1.2$~T which is accompanied by pronounced length changes. In addition, high-resolution thermal expansion data are used for detailed analysis of pronounced magneto-elastic coupling in \nto. Gr\"{u}neisen scaling of the magnetic contributions to \cp\ and \alv\ implies a single magnetic degree of freedom driving the observed length and entropy changes at \tn . Our analysis suggests in-plane ferromagnetic interactions mainly determine the value of \tn . Relating our findings to recently reported strong magnetodielectric effects in \nto\ implies the essential role of structural changes for magneto-dielectric coupling and suggests a single magnetic origin of low-temperature dielectric, structural and magnetic changes in \nto.

\begin{acknowledgements}
Partial support by Deutsche Forschungsgemeinschaft (DFG) via Project KL 1824/13-1 is gratefully acknowledged. KD acknowledges fellowship by the IMPRS-QD. SS acknowledges financial support from the Science and Engineering Research Board (SERB) under grant no. EMR/2016/003792. We acknowledge the support of the HLD at HZDR, member of the European Magnetic Field Laboratory (EMFL).
\end{acknowledgements}

\end{document}